\documentclass[twocolumn,usenatbib,useAMS]{mn2e}
\usepackage{natbib}
\usepackage{amsbsy}
\usepackage{amssymb}
\usepackage{amsmath}
\usepackage{graphicx}
\usepackage{xcolor}
\usepackage[caption = false]{subfig}
\usepackage [english]{babel}
\usepackage [autostyle, english = american]{csquotes}
\usepackage{natbib}
\MakeOuterQuote{"}

\begin{document}

\title [Crustquake initiated neutron excitation \& glitches] 
{Vortex unpinning due to crustquake initiated neutron excitation and pulsar glitches}
\author[Layek, Yadav]{Biswanath Layek$^1$ \& Pradeepkumar R. Yadav$^{1}$ \\
\\
$^1$Department of Physics, Birla Institute of Technology and Science, Pilani Campus, Pilani, 
Jhunjhunu 333031, Rajasthan, India}

\maketitle

\begin{abstract}
Pulsars undergoing crustquake release strain energy, which can be absorbed in a small region 
inside the inner crust of the star and excite the free superfluid neutrons therein. The scattering of 
these neutrons with the surrounding pinned vortices may unpin a large number of vortices and effectively reduce 
the pinning force on vortex lines. Such unpinning by neutron scattering can produce glitches for Crab like
pulsars and Vela pulsar of size in the range $\sim 10^{-8} - 10^{-7}$, and $\sim 10^{-9} - 10^{-8}$, respectively. 
Although we discuss here the crustquake initiated excitation, 
the proposal is very generic and equally applicable for any other sources, which can excite the free superfluid neutrons, 
or can be responsible for superfluid - normal phase transition of neutron superfluid in the inner crust of a pulsar. 
\end{abstract}


\section{Introduction}
Pulsars are known to be excellent time-keepers. However, a significant number of pulsars show sporadic spin-up events, namely glitches. A total of 555 glitches in 190 pulsars have been catalogued\footnote{see for updated catalogue on \\http://www.jb.man.ac.uk/pulsar/glitches/gTable.html} and reported to date \citep{espinoza}. The size of glitches lie in the range 
$\sim 10^{-12} - 10^{-5}$, with a typical interglitch time of a few years. Although, the models associated with pinning-unpinning of superfluid vortices \citep{anderson75} are considered to be the most popular models for explaining glitches, the crustquake \citep{ruderman69} model finds its place in the literature quite regularly in the study of glitches or otherwise (see \cite{haskell} for a detailed review). There have been discussions in the literature suggesting the involvement of crustquake in neutron star physics, such as an explanation for the giant magnetic flare activities observed in magnetars \citep{thompson1995,lander2015}, as a possible source of gravitational waves (GW) from isolated pulsars \citep{keer2015, pradeep}. On the other hand, though the basic picture of the superfluid model is well accepted, it has a few issues which are not understood with certainty. For instance, the value of a very important quantity used in this model, namely, the pinning strength is not calculated from first principles. Similarly, the precise mechanism which triggers vortex unpinning is not known with certainty. In fact, there are suggestions 
\citep{melatos2008, eichler2010} that crustquake itself might act as a trigger mechanism for sudden angular momentum transfer by vortices to the crust. Similarly, \cite{akbal2017} have proposed that the movement of the broken crustal plate caused due to crustquake, and hence, the motion of vortices attached with the plate is responsible for glitches. Here we should also mention an interesting study 
by \cite {link-epstein96}, where the authors have proposed thermally driven pulsar glitches, caused by 
sudden deposition of energy in the inner crust. The deposited energy propagates as thermal waves through some parts of the inner crust and  
raises the local temperature. As per their study, this thermal fluctuations caused by energy depositions affect the coupling between neutron 
superfluid and the rigid outer crust causing the star to spin-up. Crustquake has been assumed to be one of the sources for such energy depositions. 
Although our proposed mechanism of glitches is unpinning of vortices, as we discuss below, in contrast to the picture as proposed by \cite{link-epstein96}, 
the work is worth mentioning from the perspective of energy deposition into the crust caused by crustquake.

In this work, we propose a novel mechanism of unpinning of superfluid vortices. In our scenario, crustquake and superfluid vortices are both considered to be responsible for pulsar glitches. We propose that strain energy released ($\Delta E = B \Delta \epsilon$) due to crustquake \citep{baym1971} excites the unbound superfluid neutrons that exist in the inner crust. These excited neutrons should share their energy with the pinned vortices through scattering. As a result, a large number of vortices ($\sim 10^{13}$) existing in the neighbourhood of the quake site can be released, causing the star to spin-up. Here, 
the size of the glitch depends on the number of vortices released due to the excitations. For Crab like pulsars and Vela, the size of the 
glitches will be shown to lie in the range $\sim 10^{-8} - 10^{-7}$ and $\sim 10^{-9} - 10^{-8}$, respectively. \\

The paper is organized in the following manner. In section 2, we briefly review the relevant features of the crustquake model. We present our work in the subsequent
sections. In section 3, we determine the number of vortices that can be unpinned through neutron-vortex scattering. Here, we will provide the expression for 
glitch size. The mechanism for unpinning of vortices and the time of occurrence of glitches will be discussed in sections 4 and 5, respectively.  We will present our results in section 6. Here, we will also make a brief comment on the future scope of this study. Finally, we conclude our work in section 7.

\section{Crustquake :  The Basic Features}
The crustquake in a neutron star is caused due to the existence of a solid elastic  deformed crust of thickness about 
1 km \citep{ruderman69, smolu70}. The deformation of the crust is characterised by its ellipticity/oblateness $\epsilon = \frac{I_{zz} - I_{xx}}{I_{0}}$, where $I_{zz}$, $I_{xx}$ and $I_0$ are the moment of inertia about z-axis (rotation axis), x-axis and of the spherical star, respectively \citep{baym1971}. At an early stage of formation, the crust solidified with initial oblateness $\epsilon_0$ (unstrained value) at a much higher rotational frequency of the star. As the star slows down, the oblateness $\epsilon (t)$ decreases, leading to the development of strain in the crust. Once the critical stress is reached,  crust cracks followed by a sudden change of oblateness $\Delta \epsilon$. As a result, the moment of inertia (MI) of the star decreases, which leads to an increase in its rotational frequency. In the crustquake model, the glitch size is related to $\Delta\epsilon$ through $ \frac {\Delta \Omega}{\Omega} = -  \frac {\Delta I}{I_0} = \Delta \epsilon$, and it is completely determined by the extent to which $\Delta\epsilon$ is changed due to crustquake. The interglitch time being proportional to $\Delta\epsilon$, is also determined by the change of oblateness. Note that for crustquake to be a successful model for glitches, it should be a regular event for
a pulsar, with a frequency of once in a few years. This requires the crustal strain to be released partially in a crustquake event. Equivalently, 
the fractional strain release, $\eta = \frac{\Delta\epsilon}{\epsilon}$ should always be less than unity. In fact, smaller the value of this fraction, larger the number of crustquake events that are likely to occur during the rotational history of a pulsar.
In our model, we will take a fixed value of $\Delta\epsilon = 10^{-8}$ motivated by the crustquake model of glitches for Crab like pulsars. By \lq Crab like\rq, we mean pulsars with the characteristic age \citep{andersson2012} and glitch size similar to that of Crab pulsar. We will show below that the condition $\eta < 1 $ will be satisfied with the realistic values of star's 
ellipticity $\epsilon$, which is determined by the critical (breaking) strain $u_{cr}$ that a star can sustain without breaking. \\

We will now present the relationship between the critical strain $u_{cr}$ and the ellipticity of the star $\epsilon$ by mentioning a few relevant parameters (see \cite{baym1971,jones2002} for details).  The total energy of the deformed 
pulsar is given by \citep{baym1971}, 
\begin{equation} \label{eq:energy}
E = E_0 + \frac{L^2}{2I} + A\epsilon^2 + B(\epsilon- \epsilon_0)^2 .
\end{equation}
Where $E_0$ is the contribution of gravitational potential energy of the spherical pulsar. $L$ and $I$ are the angular momentum
and the moment of inertia of the deformed pulsar, respectively. The coefficient $A$ ($ \simeq 10^{53}~ \text {erg}$) arises as a correction of gravitational energy due to deviation from sphericity. The coefficient $B$ ($\simeq 10^{48} ~\text {erg}$) is related to the modulus of rigidity of the star's crust \citep{baym1971,jones2001}. Within an approximation $B << A$, the upper bound on star's ellipticity can be written in terms of the critical strain $u_{cr}$ as
\citep{baym1971,jones2002,pradeep},
\begin{equation} \label{eq:strain}
\epsilon < \frac {B}{A} u_{cr} \simeq 10^{-5} u_{cr}.
\end{equation}
Using the above equation, the possible upper bound on $\epsilon$ can be obtained from the values of $u_{cr}$, as estimated theoretical in several works \citep{horowitz2009, chugunov2010, baiko2018}. \cite{horowitz2009} have done detailed molecular
dynamics simulations to estimate the magnitude of crustal breaking strain of neutron star. Simulations were performed through modelling the crust as monocrystalline and polycrystalline materials and they obtained the critical strain $u_{cr} = 0.1$. Substituting this in Eq. (\ref{eq:strain}), we get $\epsilon < 10^{-6}$. 
Recently, \cite{baiko2018} have followed a semi-analytical approach to study the crustal properties of a neutron star, including the analysis on crustal breaking strain. For polycrystalline 
materials, they have obtained the value $u_{cr} = 0.04$. For this value, the upper bound of ellipticity is given by $\epsilon = 0.4 \times 10^{-6}$. From the observational perspective, there were several attempts \citep{vela2006, aasi2013, aasi2014, ligo2020} to constrain star's ellipticity by observations. Any asymmetric mass distribution of pulsar relative to its rotation axis, such as triaxiality \citep{jones2002}, mountains \citep{haskell2006mountain, sudip2020} that can be characterised by ellipticity parameter are the source of continuous gravitational waves. As the strain amplitude of such gravitational waves is proportional to $\epsilon$, their non-observation naturally puts an upper limit on the ellipticity of the star. Here, we mention recent results by \cite{ligo2020}, that are based on the analysis of LIGO and VIRGO data obtained from the searches of 
continuous gravitational waves from a few selected isolated pulsars. The results were presented for three recycled pulsars,  
along with two relatively young pulsars Crab and Vela. We will quote the results for Crab and Vela that are relevant 
in the context of our present model. As per the analysis in \cite{ligo2020}, the upper limits of $\epsilon$ were
constrained at $10^{-5}$ and $10^{-4}$ for Crab and Vela, respectively. \\

The typical fractional strain released $\eta = \frac{\Delta \epsilon}{\epsilon}$  can now be obtained for 
a fixed value of $\Delta \epsilon = 10^{-8}$ and 
putting the values of $\epsilon$ as quoted above. Firstly, within the theoretical uncertainties in the estimate of $u_{cr}$, 
the values of $\epsilon$ in the range $(1.0 - 0.4) \times 10^{-6}$ provides $\eta$ in the range $\sim ~ 0.01 - 0.02$. For the values of $\epsilon$ as constrained by the observations, $\eta$ will be even smaller. For Crab and Vela, the values are given by $\eta = 10^{-3}$ and $\eta = 10^{-4}$, respectively. As we see from above, the set of values of $\eta$ satisfy the condition 
$\eta < 1$ quite comfortably. Hence, we will take $\Delta\epsilon = 10^{-8}$ throughout this work to be consistent with the conditions that are
required in the crustquake model, i.e., the glitch size, interglitch time and the fractional strain released. The corresponding value of strain energy is then
given by
$\Delta E = B \Delta \epsilon \simeq 10^{40}$ erg. We assume that the released energy is absorbed in the inner crust and excites the neutrons in the bulk neutron superfluid. We will show that the excited neutrons, in turn, can unpin a large number of vortices through neutron-vortex scattering from a local region in the 
equatorial plane. We 
calculate the number of unpinned vortices and estimate the glitch size. Note, with the fixed value of $\Delta\epsilon = 10^{-8}$, the interglitch time is always fixed to be about one year, irrespective of the glitch size produced by the local unpinning in our model. We will show that for Crab like pulsars and Vela, the glitch size still can vary in the range $\sim 10^{-9} - 10^{-7}$, without affecting the interglitch time. \\

\section {\bf {Excitation of Superfluid Neutrons and Glitches}}
\begin{figure}
\centering
\vspace{0.5cm}
\includegraphics[width=1.0\linewidth]{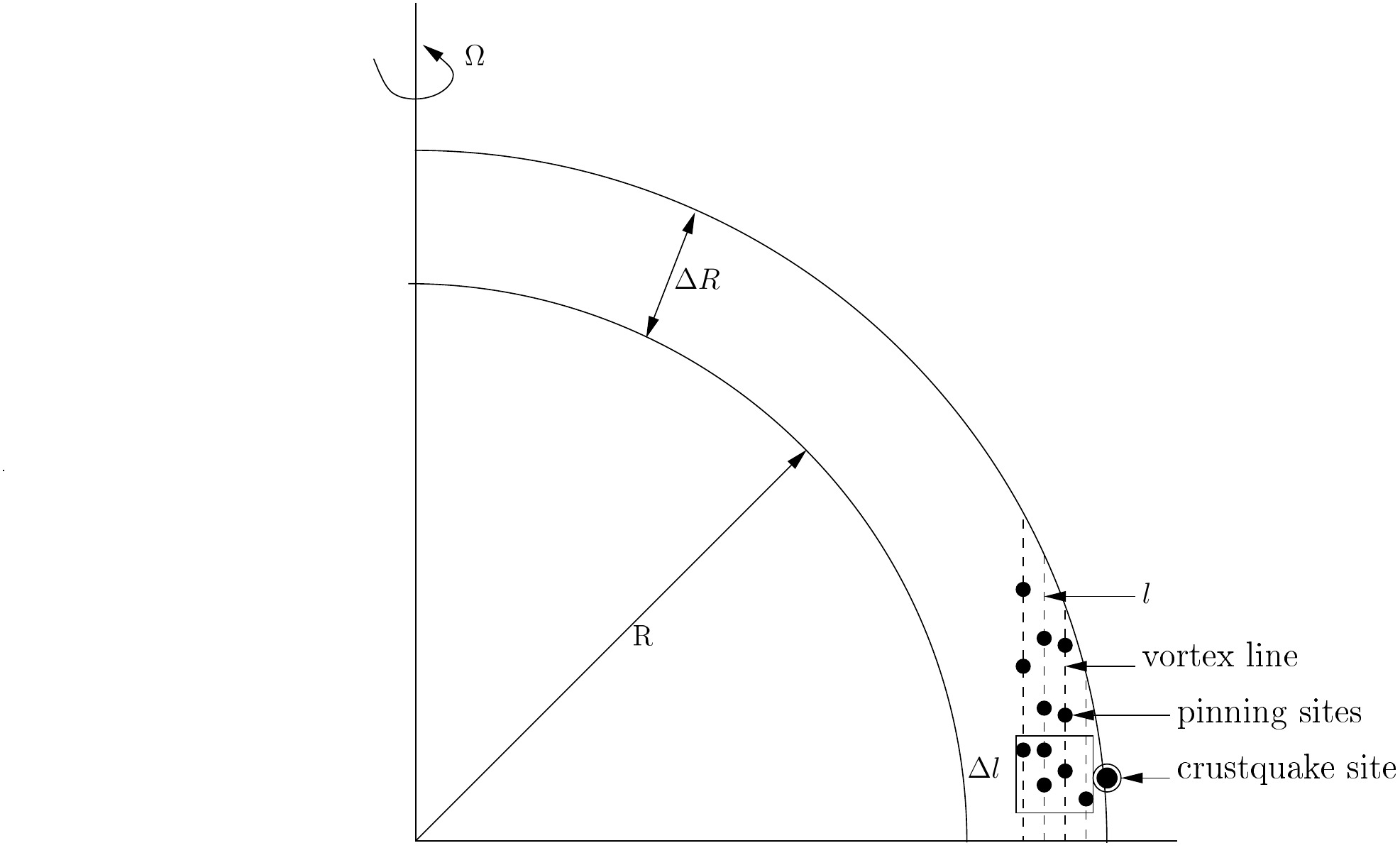}
\caption{Schematic representation (cross-sectional view, not to scale) of quake site and the nearby volume element $V_p$, where local unpinning may occur by 
absorbing energy due to crustquake. $R$ ($\simeq 10$ km) is the distance of the inner crust from the centre of the star and $\Delta R \simeq 1$ km is the 
thickness of the crust.}
\label{fig:geometry}
\end{figure}

We assume that superfluid vortices are pinned at $t = 0$, and a fraction of these vortices get unpinned by neutron excitations caused by crustquake at $t = t_p$. 
The interglitch time $t_p$ is expected to be of the same order as the time duration of successive crustquake events. 
We take the picture that crustquake occurs in a local region around the equatorial plane in the outer crust of the star (see Fig. \ref{fig:geometry}). We choose the quake site to be in the equatorial plane motivated by the picture
proposed by \cite{baym1971} in their work on the crustquake model for glitches. There was also a detailed study \citep{franco2000} on the development of the crustal strain, which arises due to the slowing down of the star. By including the effects of the magnetic field, the authors have calculated the strain angle and found out that the strain angle is maximum in the equatorial plane, making it most likely place for the quake site. \\

 The absorption of strain energy $\Delta E = B \Delta \epsilon$ should excite 
the free superfluid neutrons that exist outside the nuclei surrounding the pinned vortices. The sharing of energy through scattering by these 
excited neutrons with the vortex core neutrons should result in the unpinning of vortices causing the glitch event. 
Assume $\Omega_p$ is the angular velocity of the pinned vortices that remains fixed during $t = 0$ to $t = t_p$. $\Omega_c (t)$ is the angular velocity of the
corotating crust-core coupled system with $\Omega_c (0) = \Omega_p$. The development of differential angular velocity $\delta \Omega = \Omega_p -\Omega_c (t)$
between vortices and the rest follows the time evolution of the star and can be written as (at $t = t_p$)
\begin{equation} \label{eq:domega1}
\frac{\Omega_p - \Omega_c (t_p)}{\Omega_c (t_p)} \equiv \Big(\frac {\delta \Omega} {\Omega}\Big)_{t_p} \simeq \frac {2~t_p}{\tau}, 
\end{equation}
where $\tau = - \frac{\Omega}{2 \dot \Omega} $ is the characteristic age of pulsar and we assume $t_p << \tau $. For ease of notation, from now onward we assume, $(\frac {\delta \Omega} {\Omega})_{t_p} 
\equiv \frac {\delta \Omega} {\Omega} $. The glitch size can be estimated applying the model of superfluid vortices,
\begin{equation} \label{eq:Domega1}
\frac{\Delta \Omega}{\Omega} = \Big(\frac{I_p}{I_c}\Big) \Big(\frac{\delta \Omega}{\Omega}\Big) \Big(\frac{N_v}{N_{vt}}\Big),
\end{equation}
where $\frac{I_p}{I_c}$ is the MI ratio of bulk neutron superfluid in the inner crust to the rest of the star \citep{ruderman1976}. $N_{vt}$ is the total number of pinned vortices in the equatorial plane in the inner crust. The ratio $\frac{N_v}{N_{vt}}$ is incorporated since only a fraction of vortices is expected to 
be affected by the excited neutrons. In standard superfluid model \citep{anderson75}, the above ratio is almost unity and $\delta \Omega$ should be replaced by 
its critical value $\delta \Omega_{cr}$. Where $\delta \Omega_{cr}$ is the maximum value of $\delta \Omega$ at which 
the magnus force balances the pinning force. The magnus force per unit length on a vortex line is given by $f_m = \rho \kappa R \delta \Omega$. Equating this with the pinning force per unit length
$f_p = \frac{E_p}{b \xi}$,  we get \citep{alpar1984},
\begin{equation} \label{eq:domegacr}
\delta \Omega_{cr}= \frac{E_p}{\rho \kappa R b \xi},
\end{equation}
where, $E_p$ and $\rho$ are the pinning energy per pinning site (to be estimated later) and the local mass density, respectively.  $\kappa = \frac{h}{2m_n}$ is the quantum vorticity with $m_n$ being the neutron mass. $\xi \simeq 10 ~\text{fm}$ is the coherence length of the bulk superfluid, and the nucleus-nucleus distance is denoted by $b~ (\simeq 100~\text{fm})$. $R$ ($\simeq 10$ km) is the distance of the inner crust from the centre of the star. The numerical value of $\delta \Omega_{cr}$ will be estimated in the next section.
We now estimate the number of vortices $N_v$ that are expected to be affected by the neutron excitation. First, we take a region of volume $V_p$ within which the free neutrons should be excited. The relevant volume can be estimated by energy balance as 
\begin{equation} \label{eq:energy}
B \Delta \epsilon = N_e \Delta_f = \frac{\Delta_f^2}{E_f} n_f V_p
\end{equation}
or equivalently,  
\begin{equation} \label{eq:vp1}
V_p = \frac{B \Delta \epsilon ~E_f}{n_f \Delta_f^2} ,
\end{equation}
where the free neutron number density in the superfluid state is denoted by $n_f$, 
and $\Delta_f$ denotes the superfluid free energy gap of the neutrons. $N_e$ is the number of excited neutrons. 
The mass density $\rho$ of the inner crust lies in the range $\sim (10^{11} - 10^{14})~\text{gm-cm}^{-3}$.
The Fermi momentum of the free neutrons at $\rho \simeq 5 \times 10^{11} ~\text{gm-cm}^{-3}$ has been calculated by several authors \citep{pastore2011, chamel2008, sinha2015} and found to be of order $k_f \simeq 0.20~\text{fm}^{-1}$. The corresponding value of neutron density is given by, $n_f = \frac{k_f^3}{3 \pi^2} = 2.7 \times 10^{-4}~\text{fm}^{-3}$, which increases as one goes deeper in the crust. For our case, the relevant region of interest is the outer part of the inner crust, and we will take the above value for the estimate of $V_p$. The superfluid gap parameter $\Delta_f = 0.06 ~\text {MeV}$ for $k_f \simeq 0.20~\text{fm}^{-1}$ \citep{pastore2011, chamel2008, sinha2015}. Putting the values of various quantities in Eq. (\ref{eq:vp1}), we get $V_p = 5.1 \times 10^6 ~\text{m}^3$.
Note that the calculation of $V_p$ assumes the isotropic distribution of energy from the quake site, and the geometry is taken 
to be a cubical shape as shown in Fig. \ref{fig:geometry}.\\

The isotropic distribution of energy from the quake site is an assumption in estimating the volume of the affected region. 
In principle, the volume of the affected region should depend on the Fermi energy ($E_f$) and pairing energy 
($\Delta_f$) of the neutron superfluid in that region. 
Although, the distances to which the energy transports in the azimuthal direction (with respect to the rotation axis) and 
along the altitude of the star are expected to be the same, the distance across the inner crust should be different from the other
two directions. In this work, we will not take into account such anisotropy and assume
cubical geometry only. Here, we mention the works of \cite{lander2015} and \cite{akbal2017}, where the authors have considered
a cubical geometry in their respective studies. \cite{lander2015} have studied the crustquake event due to the development of magnetic strain as a result of internal magnetic field evolution of the star. In their study, the authors have considered 
a cubical crustquake geometry to calculate the required magnetic field strength for crust breaking. 
Similarly, \cite{akbal2017} have studied vortex unpinning (Although, the unpinning mechanism is completely different from ours.) 
due to the crustal plate movement triggered by crustquake. In their work, the size of the broken plate and the number of unpinned vortices are calculated by modelling cubical shape as one of the quake site geometries. The number of unpinned vortices estimated 
by the authors turned out to be of a similar order, irrespective of the assumed geometries in their work. In this spirit and for simplifying calculations, we consider a cubical geometry to test our model by estimating the volume $V_p$ of the affected region, number of unpinned vortices $N_v$ and the glitch size $\frac{\Delta \Omega}{\Omega}$. \\

Now denoting $\Delta l$ as the length of each side of the cube, we now express the volume $V_p$ in terms of number of vortices $N_v$ as  
\begin{equation} \label{eq:vp2}
V_p = (\Delta l)^3 = \frac {N_v \Delta l}{n_v},
\end{equation}
where $n_v = \frac{2m_n \Omega}{\pi \hbar} = 10^3 ~\text{cm}^{-2} (\frac {\Omega}{\text{s}^{-1}})$ is the areal density of vortices. Number 
of vortices in the equatorial plane, which are expected to be unpinned due to neutron excitation, can be estimated using Eq. (\ref{eq:vp1})
and Eq. (\ref{eq:vp2}) as
\begin{equation} \label{eq:nv}
N_v = \frac{B \Delta \epsilon ~n_v ~E_f}{\Delta l  ~n_f \Delta_f^2} = 3.1 \times 10^{11}~\Big (\frac {\Omega}{\text{s}^{-1}}\Big). 
\end{equation}
In the above equation, the numerical factor has been calculated by taking   
$\Delta l = (V_p)^{1/3} = 172$ m and the values of other parameters are taken as mentioned earlier. Note that for Crab/Vela $\Omega \simeq 10^2$, 
there are about $N_v = 10^{13}$ vortices which can be released from the volume $V_p$. Finally, using Eq. (\ref{eq:domega1}) and Eq. (\ref{eq:Domega1}), the glitch size is obtained as 
\begin{equation} \label{eq:Domega2}
\frac{\Delta \Omega}{\Omega} = \Big (\frac{6.2 \times 10^{11}}{N_{vt}}\Big) \Big (\frac {\Omega}{\text{s}^{-1}}\Big) 
\Big(\frac{I_p}{I_c}\Big) \Big(\frac{t_p}{\tau}\Big),
\end{equation}
where the total number of vortices in the crust is given by 
\begin{equation} \label{eq:nvt}
N_{vt} \simeq (2 \pi R \Delta R) n_v = \Big (\frac{6.3 \times 10^{14} \Omega} {\text{s}^{-1}}\Big) .
\end{equation}
Here, $\Delta R \simeq 1$ km is the thickness of the crust. Substituting the value of $N_{vt}$ in Eq. (\ref{eq:Domega2}), we get
\begin{equation} \label{eq:Domega3}
\frac {\Delta \Omega}{\Omega} \simeq 10^{-3}\Big(\frac{I_p}{I_c}\Big) \Big(\frac{t_p}{\tau}\Big).
\end{equation}
By taking $t_p$ of the same order as typically observed interglitch time of pulsars, we estimate the glitch 
size using Eq. (\ref{eq:Domega3}) and results are discussed in section 6. Of course, the glitch will arise 
provided $N_v \simeq 10^{13}$ vortices are unpinned by neutron excitation from the region of our interest. We now 
discuss the mechanism which ensures the unpinning of the vortices.

\section {Mechanism of Local Unpinning}
We assumed that the crustquake occurs in the outer crust in the vicinity of the equatorial plane and the energy released by this event is 
distributed isotropically from the quake site. The energy $\Delta E \simeq 10^{40}$ erg absorbed in the volume 
$V_p = 5.1 \times 10^6~\text{m}^3$ should excite a $N_e$ ($ = \frac{\Delta}{E_f} n_f V_p \simeq 10^{47}$) number of neutrons from the bulk neutron superfluid. Ignoring small finite temperature ($kT \simeq 0.01~\text{MeV}$) correction, each of these excited neutrons has approximately  
$E_f = 0.83~\text{MeV} $ amount of energy. If the total energy of the excited neutrons is more than the pinning energy of all vortices enclosed in volume $V_p$ (see Fig. \ref{fig:geometry}), then the inelastic scattering of these neutrons with the vortex core neutrons should unpin the vortices. Note that the pinning energy $E_p$ acts as the binding energy of the vortex-nucleus system, and it arises due to the interaction of the vortex with the nucleus. Thus the sharing of energy by the excited neutrons with the vortex core neutrons increases the kinetic energy of the later. In fact, the energy of the excited neutrons equivalently can be treated
as the activation energy, which helps to overcome the pinning barrier. We will show below that the excited neutrons have the required energy to overcome the barrier. The inelastic collision can be represented as,\\

{\it excited neutron ($\sim E_f$) + pinned vortex ($ - E_P$) $\rightarrow$ de-excited neutron ($E_f - E_p$) + 
free vortex.} \\

The quantities in bracket denote the energy of various objects. The negative sign in front of $E_p (>0) $ signifies the binding energy of the pinned vortex. It should be noted that following unpinning; the vortex is free to move (i.e., free vortex) outward with radial velocity $v_r$. Here we make a few comments on the possibility of repinning of the outward moving vortices. Since the study of repinning initiated by \cite{sedrakian1995}, the mechanism and consequences of repinning have been discussed often in the literature, whether in the context of creep theory \citep{alpar1984}, or in the standard theory of superfluid vortices \citep{anderson75}. One of the important consequences of repinning, namely, the acoustic radiation is quite relevant to our present 
model. As per the studies in \cite{warszawski2012, warzawski2013}, the
acoustic radiation caused by repinning (so-called \lq acoustic knock-on\rq $~$as per the terminology used in \cite{warszawski2012}) is believed to play an
important role in the process of vortex avalanche. The avalanche can be a viable process for our model to produce large size glitches through acoustic 
knock-on (and proximity knock-on) caused by repinning of outgoing vortices. In this present work, we skip the
studies of vortex avalanche (which will be explored in future) except making a few comments in the 
results \& discussion section. \\

 We now compare the pinning energy with the energy of excited neutrons. The pinning mechanism as per discussion in Ref. \cite{alpar1984} depends on the density $\rho$. It was suggested that for $\rho > 10^{13} \text{gm-cm}^{-3}$, the vortex lines are pinned 
to lattice nuclei with pinning energy per site \citep{alpar1984,link1991},
\begin{equation} \label{eq:ep}
E_p = \frac{3}{8} \gamma \frac{\Delta^2}{E_f} n_f V \simeq 5.8 \times 10^{-4} ~ \text{MeV}.
\end{equation}
Where $V  = \frac{4}{3} \pi \xi^3$ is the overlap volume between the vortex and the nucleus. The size of vortex core $\xi$ ($\simeq 10 ~\text {fm}$) is of the same order as the nuclear radius. The numerical value of $\gamma$ is of order unity \citep{alpar1984, alpar1989vortex}. For the density $\rho < 10^{13} \text{gm-cm}^{-3}$, the lines are preferably pinned in 
between nuclei (interstitial pinning) and the pinning energy $E_p$ has been calculated to be of the order 1 KeV per site \citep{link1991}, 
which is approximately the same as obtained in Eq. (\ref{eq:ep}). The above two pictures though qualitatively different, provide approximately the same value of $E_p$ for the pinning region of our interest and both pictures are consistent with our unpinning mechanism. The total amount of pinning energy associated with all the vortices within 
volume $V_p$ can be estimated as
\begin{equation} \label{eq:etp}
E_{tp} = \Big(\frac{E_p}{d_v^2 b}\Big) V_p = 3.1 \times 10^{24}  ~\text{MeV}.
\end{equation}
Where $d_v$ ($ = 0.01 ~\text{cm} $) and $b$ ($= 100 ~ \text{fm}$) are inter vortex and inter-nuclear 
distance, respectively. The value of $E_p$ is taken from Eq. (\ref{eq:ep}). The energy as estimated in Eq. (\ref{eq:etp}) should now be 
compared with the total kinetic energy of the excited neutrons in the same volume $V_p$ ,
\begin{equation} \label{eq:ete}
E_{te} \simeq N_e E_f = \Big(\frac{\Delta}{E_f} n_f\Big) E_f V_p  = \Delta ~ n_f V_p \simeq 10^{47} ~ \text{MeV}.
\end{equation}
Thus, we see from Eq. (\ref{eq:etp}) and (\ref{eq:ete}) that the excited neutrons have the required energy to unpin the vortices
that are contained in the volume $V_p$.\\

For the sake of numbers, with a typical inter-nuclear distance of about a 100 fm at baryon density $\rho \simeq 10^{11}~\text{gm-cm}^{-3}$, 
a single vortex line passes through $ \frac{\Delta l}{b} \simeq 10^{15}$ number of nuclei  within the cube of 
height $\Delta l \simeq 172 $ m and about $10^{13}$ nuclear sites which host pinned vortices in a single vortex line will be affected due to 
the neutron-vortex scattering. Since all the vortices that lie inside the volume $V_p$ are unpinned, the pinning force on the whole vortex lines 
passing through $V_p$ is reduced. The fractional decrease in pinning force per unit length is given by
\begin{equation} \label{eq:force}
\frac {\Delta f_l}{f_l} = \frac {\Delta l}{l} \simeq 0.1,
\end{equation}
where $f_l$ is the pinning force per unit length \citep{anderson75, pizzochero2011} of vortex lines and $\Delta f_l$ is the decrease of 
pinning force per unit length due to the effects as mentioned above.  
The above numerical value in  Eq. (\ref{eq:force}) has been calculated by taking the length of a vortex line (assuming straight), threaded in 
the inner crust (see Fig. \ref{fig:geometry}) as, $l \simeq \sqrt {2 R \Delta l} \simeq 1855 $ m ( R = 10 km).  
The decrease in pinning force per unit length for a vortex line must reduce the critical differential angular frequency, 
$\delta \Omega_{cr}$. The numerical value of $\delta \Omega_{cr}$ can be estimated from Eq. (\ref{eq:domegacr}). Taking
$E_p = 5.8 \times 10^{-4}~\text{MeV}$ from Eq. (\ref{eq:ep}) and  $\rho = 5 \times 10^{11}~\text{gm-cm}^{-3}$, we get
$\delta \Omega_{cr} = 0.09~\text{rad-s}^{-1}$. The fractional decrease of the above quantity should be of the same order as 
$\frac {\Delta f_l}{f_l}$,  i.e., 
\begin{equation} \label{eq:omegac}
\frac {\Delta (\delta \Omega_{cr})}{\delta \Omega_{cr}}  \simeq 0.1 .
\end{equation}
If the build-up differential angular frequency $\delta \Omega$ at $t_p$ does not differ much from its critical value 
$\delta \Omega_{cr}$, then the magnus force effectively should be able to move the vortex lines from the pinning site. 
The relative difference can be estimated as 
\begin{equation} \label{eq:diff}
\frac {\delta \Omega_{cr} - \delta \Omega} {\delta \Omega_{cr}} = 1 - \Big(\frac {2 t_p}{\tau}\Big) \Big(\frac{\Omega}{0.09 \text{s}^{-1}}\Big) 
\equiv 1 - x .
\end{equation}
The numerical value of $(1 - x)$ can be determined by taking the typical values for Crab/Vela, 
$t_p = 1 ~\text{year}$, $\tau = (10^3 - 10^4) ~ \text {years}$ and $\Omega \simeq 10^2 ~\text{s}^{-1}$. For
these set of values, the numerical value of $(1 - x)$ turns out to be of a similar order, such 
that Eq. (\ref{eq:omegac}) is satisfied. Note, the purpose of the above exercise is to check whether the pinning force is 
decreased enough for the vortex lines passing through
the volume  $V_p$ to move under magnus force, even when $\delta \Omega$ at $t = t_p$ is less than the critical value 
$\delta \Omega_{cr}$. It is indeed true, as suggested by the above set of arguments. We conclude this part
by saying that crustquake initiated neutron excitations can unpin a large number of vortex lines which pass through the volume $V_p$ and hence produces glitch through the local unpinning.\\
\section{The time of occurrence of glitches following crustquake}
Now we estimate the time of occurrence of glitches $t_g$ following the crustquake event. The time $t_g$ is 
the sum of the time taken for unpinning followed by the time for the vortex to move outward and share their excess angular momentum to 
the crust. First, we estimate the time for unpinning, which is determined by the relaxation time scale ($\tau_{nn}$) of neutron-vortex scattering. 
The calculation of $\tau_{nn}$ deserves a separate work, and we will provide here an order of magnitude estimate following the approach of \cite{feibelman1971}. 
To understand postglitch behaviour of pulsars, the author \citep{feibelman1971} has worked out a detailed calculation of relaxation time scale 
$\tau_{en}$ of electrons by considering the scattering of thermal electrons with the vortex core neutrons. The contribution of neutron-vortex 
scattering was not considered in their calculation. It is mainly due to the presence of very few thermally excited neutrons in the bulk superfluid at a lower temperature ($kT \simeq 0.01$ MeV) of the star. Note that the probability of excited neutrons in the bulk superfluid is suppressed by the factor $e^{-\frac{\Delta_f}{kT}}$.
 Hence, the thermal neutron-vortex scattering is expected to be negligible compared to electron-neutron scattering. 
However, the presence of excited neutrons in our case is not of thermal origin. The neutrons are excited by the absorption of energy $B \Delta \epsilon$, and the fraction of excited neutrons relative to superfluid neutrons is given by  $\frac{\Delta_f}{E_f} \simeq 0.07$. Therefore, the scattering of these neutrons with the 
vortex core neutrons should not be suppressed, and it should naturally provide a finite relaxation time scale $\tau_{nn}$.\\

We will now estimate $\tau_{nn}$ following the expression of $\tau_{en}$ (see Equation (45) of \citep{feibelman1971}),
\begin{multline} \label{eq:tau1}
\tau_{nn} = \Big(\frac {\Omega_{c2}}{\Omega}\Big) \Big(\frac{4}{3\pi^2 g_n^4}\Big) \Big(\frac{E_f}{E_{fv}}\Big)^2 \Big(\frac{E_{fv}}{\Delta_v}\Big) \Big(\frac{E_{fv}}{2m_nc^2}\Big)^{1/2} \times \\
\Big(\frac{\hbar}{\Delta_v}\Big) \Bigg(\frac {exp \Big(\frac{1}{\sqrt{4\pi}}\Big)} {K_0\Big(\frac{1}{\sqrt{4\pi}}\Big)}\Bigg) 
exp\Big(\frac{\pi \Delta_v^2}{4\Delta_f E_{fv}}\Big)~ \text{s} .
\end{multline}
Here, $\frac {\Omega_{c2}}{\Omega}$ is the ratio of upper critical angular speed of neutron fluid ($\Omega_{c2} = 10^{20}$) 
to the angular speed of the star. For Crab/Vela the ratio is of order $10^{18}$. The coupling strength associated with neutron-neutron interactions can
be described by the neutron $g$ factor, $g_n = -1.91$. $K_0(\frac{1}{\sqrt{4\pi}})$ is the zero order Bessel
function. $E_{fv}$ and $\Delta_v$ denote the Fermi energy and superfluid gap parameter associated with the neutron vortex core, respectively. The factor 
$\beta = \frac {1}{kT}$ in the calculation of $\tau_{en}$ \citep{feibelman1971} is due to the finite temperature 
probability distribution of electrons. In our case, $\beta$ should be replaced by bulk superfluid energy gap $\Delta_f$.
For the case of nuclear-vortex pinning, we take the approximation $E_{fv} \simeq E_f = 0.83$ MeV and $\Delta_{fv} \simeq \Delta_f = 0.06$ MeV. 
For interstitial pinning, the approximation will be replaced by equality. Substituting these quantities in Eq. (\ref{eq:tau1}), we get
\begin{equation} \label{eq:tau2}
\tau_{nn} = 3.0 \times 10^{-5} \text{s} .
\end{equation}
Thus, as we see from Eq. (\ref{eq:tau2}) that the unpinning of vortices are almost instantaneous. Note, the relaxation time scale
for electron $\tau_{en}$ has been estimated \citep{feibelman1971} to be on the order of days to years. The shorter time scale of $\tau_{nn}$ as compared to 
$\tau_{en}$ is expected due to the difference of coupling strength in electron-neutron and neutron-neutron interactions. The former is dipole-magnetic moment 
interaction and hence the strength of the interaction is proportional to $ \alpha g_n$, where $\alpha = \frac{e^2}{4\pi} = \frac{1}{137}$. The later interaction 
is solely due to magnetic moment of the neutrons with the interaction strength proportional to $g_n^2$. We should mention that a detailed calculation is required for the precise estimate of $\tau_{nn}$. From the perspective of the occurrence of glitches following crustquake, as we see below that even a few orders of magnitude change in the value of $\tau_{nn}$ will have negligible contribution to $t_g$. Next, we determine the time ($t_c$) taken by the unpinned vortices to move toward the outer
crust and share their excess angular momentum to the crust. The radial velocity $v_r$ of the unpinned vortices is $v_r \simeq R (\delta \Omega) = (\frac{2 t_p}{\tau})~ \Omega R \simeq (10^5 - 10^4)~ \text{cm-s}^{-1}$, where we have used Eq. (\ref{eq:domega1}) and $\tau \simeq (10^3 - 10^4)$ years are the age of Crab and Vela, respectively.
Thus the time $t_c \simeq \frac{v_r}{\Delta l}$ lies in the range $\sim (0.17 - 1.7)~\text{s}$.  We see that the glitch due to vortex unpinning occurs at $t_g = \tau_{nn} + t_c \simeq (0.17 - 1.7)~\text{s}$ after the crustquake. The implication of the time of occurrence $t_g$ of pulsar glitches in our model will be discussed in the next section. \\

Note that the change of oblateness of the star due to crustquake is also expected to produce a glitch (as per the crustquake model) 
of order $10^{-8}$. The glitch rise time 
$\Delta t$ is approximately determined 
\citep{ruderman1991c, haskell2015} by the speed of shear wave $v = \sqrt{\frac {\mu}{\rho}} = 3 \times 10^8 ~\text {cm~s}^{-1}$. Where $\mu = 10^{30}~ \text {dynes-cm}^2$ is the shear modulus of the crust and $\rho \simeq 10^{13}~\text {gm-cm}^{-3}$ is the average crust density. Thus, the time for the shear wave to propagate along the stellar surface of the radius ($R$) is given by $\Delta t \simeq \pi R/v = 0.01~ \text {s}$ \citep{baym1971}. From the perspective of distinguishability of the glitches produced by two different sources, 
the glitch rise time $\Delta t$ in the crustquake model needs to be compared with the time of occurrence of glitch $t_g$ produced by local unpinning. We will discuss this issue in the next section.

\section{Results and Discussion}
We have discussed (in section 4) our novel mechanism of local unpinning of vortices in a given region caused by sharing of energy
by the excited neutrons with the vortex core neutrons. We will now estimate the glitch size $\frac{\Delta \Omega}{\Omega}$ caused by the 
local unpinning for Crab like pulsars and Vela pulsars using the 
Eq. (\ref{eq:Domega3}). The glitch size depends on the number of vortices $N_v$ released due to local unpinning. For
the fixed input energy $B \Delta \epsilon$, this number depends on the properties of bulk neutron superfluid via $E_f$ and $\Delta_f$.
The values of $E_f$ and $\Delta_f$ are taken from the literatures as noted in
section 3. The energy input is provided by the strain energy released $B \Delta \epsilon$ due to the crustquake. 
The value of this energy ($\simeq 10^{40}~ \text{erg}$) is set based on the arguments provided in section 2. 
The interglitch time $t_p$ is set by the frequency of occurrence of crustquake events and is proportional to the change of oblateness 
$\Delta \epsilon$ due to crustquake. We choose $\Delta \epsilon = 10^{-8}$ in accordance with the interglitch time of approximately
one year. The ratio of moment of inertia of superfluid component in the inner crust to the rest of the star is of order
$\frac{I_p}{I_c} \simeq 10^{-2}$ as per the evidences through several studies (see, for example, Ref. \cite{ruderman1976}). 
Putting these values in  Eq. (\ref{eq:Domega3}) and by taking the typical characteristic age of Crab and Vela in the range $\tau \simeq (10^3 - 10^4)$ years, we get the glitch size ($t_p \simeq 1$ year) as 

\begin{equation} \label{eq:micro}
\Big (\frac {\Delta \Omega}{\Omega}\Big) \simeq 10^{-8} - 10^{-9} ,
\end{equation}
where the relatively larger (smaller) value of glitch size is for Crab (Vela) pulsar. \\

The above estimate of glitch size corresponds to 
unpinning of $N_v \simeq 10^{13}$ vortices (Eq. (\ref{eq:nv})), out of total $N_{vt} \sim 10^{17}$ vortices (Eq. (\ref{eq:nvt})) present in
the inner crust. Note that the MI ratio $\frac{I_p}{I_c}$ is taken as $10^{-2}$ for the estimate of glitch size. It was suggested 
\citep{ruderman1976} that this ratio takes different values depending on the presence or absence of normal neutron fluid 
(called as \lq transition region\rq) between the inner crust and the interior neutron superfluid. 
In the presence of a normal layer, the unpinned vortices are required to share their excess angular momentum to a relatively
larger part of the corotating system and hence, $\frac{I_p}{I_c} \simeq 10^{-2} $ 
is relatively smaller. In the absence of such layer, the above ratio is increased and approximately is given by $\frac{I_p}{I_c} = 0.1$. In fact, \cite{piekarewicz2014} suggested that within theoretical uncertainties in the equation of state, the 
neutron star can have $\frac{I_p}{I_c} = 0.1$.
For this value, there will be about one order of magnitude enhancement in the glitch size. For Crab like pulsars and Vela, the 
glitch size can be of order $10^{-7}$ and $10^{-8}$, respectively. Interestingly, such glitches have been observed
for Crab \citep{basu2020,alpar2020} and Vela pulsar \citep{cordes1988,jankowski2015}. 
Note, the interglitch time remains same in our model due to the fixed value of $\Delta \epsilon = 10^{-8}$, even
though the glitch size varies. Thus, the correlation between the glitch size and the waiting time does not
exist in our model, which is consistent with the statistical study for most of the pulsar glitches \citep{warzawski2013}. \\

From the perspective of the time of occurrence of glitches, we mentioned that the time interval between crustquake and
the glitch produced by unpinned vortices should be decided by the relaxation time scale ($\tau_{nn}$) of 
excited neutrons, followed by the time $t_c$ taken by the vortex to reach the outer crust and share their excess angular momentum.
In section 5, we have attempted to provide an order of magnitude estimate of $\tau_{nn} (\simeq 10^{-5}~ \text{s})$ following the 
work of \citep{feibelman1971}. The total time duration turns out to be $t_g = \tau_{nn} + t_c \simeq (0.17 - 1.7)~\text{s}$, which is the 
time of occurrence of glitch by vortex unpinning after crustquake. We point out here that too small value of neutron relaxation time scale 
$\tau_{nn}$ compared to electron-neutron relaxation time scale $\tau_{en}$ is quite significant from the perspective of observation of glitches 
following crustquake. The longer time scale (few days to few months) of $\tau_{en}$ is crucial in explaining postglitch behaviour of the star. 
In contrast, $\tau_{nn}$ and $t_c$ set the time of occurrence of glitch immediately after the crustquake event. As crustquake itself produces glitch (because of rearrangement of the shape of the pulsar), the time scale of $\tau_{nn}$ of similar order as $\tau_{en}$ could have been noticeable through the recurrence of another glitch within a few days/months. The non-observation of another glitch within such interval follows from the fact
that $\tau_{nn} ~(\simeq 10^{-5}~\text{s})$ is too small. Also for the same reason, the time $t_g (\simeq t_c) $ 
can now be identified as the glitch rise time in our model, where the beginning of glitch coincides with the simultaneous 
unpinning of all the vortices. Note that $t_c \simeq  (0.2 - 2~\text{s}$) turns out to be consistent with the observation of
pulsar glitch profile (almost \lq sudden\rq spin-up event).\\

Now, if we compare $t_g$ with the glitch rise time $\Delta t \simeq 0.01~ \text{s}$ in the crustquake model, we see that the glitch produced through vortex unpinning lags behind the glitch produced due to crustquake. The conclusion is of course within the uncertainty in the estimation of $t_g$ and $\Delta t$. This 
distinguishing feature should be reflected in the pulse profile for glitches, provided the subsecond resolution in pulsar timing is achieved. 
As of now, the best resolved time observed for spin-up of Vela pulsar has been reported \citep{mcculloch1990} to be $\approx$ 2 min. In a recent work, \cite{ashton2019} have constrained the upper limit of glitch rise time for Vela pulsar to be $12.6~\text{s}$. Thus, at present, it is not possible to resolve
the glitch profile to distinguish the source of glitches. There are a few other models for explaining glitches through vortex unpinning, where
crustquake acts as a trigger mechanism (see for example \cite{ruderman1991c, link-epstein96, akbal2017}). The glitch characteristics as proposed across
these models are similar (i.e., glitch size and glitch rise time) to the predictions in our model. These common features and the constraint
in pulsar timing resolution make it difficult to isolate the precise source for glitches. \\

Before concluding the section, we make a few comments on the immediate consequences of this model, which will be explored
in our future work. First, we anticipate that 
almost instantaneous release of about $10^{13}$ vortices may act as a trigger mechanism to unpin the nearby vortices, which
lie in the equatorial plane. Among various suggestions on vortex avalanches \citep{melatos2008, warszawski2012, warzawski2013, akbal2017}, we find that the {\it knock-on} pictures \citep{melatos2008, warzawski2013} fit quite well in our model. 
In proximity knock-on, presence of the azimuthal component of the vortex velocity can make
an individual vortex knock-on the other vortices present in the equatorial plane and nearby the cubical volume $V_p$. Note that for completely outward motion, the vortices should not encounter vortices along its trajectory. In our future work, we would like 
to determine the trajectory of unpinned vortices, estimate the number of vortices that can be released through this
process. The acoustic knock-on caused by repinning of vortices can also be a viable process in our model. 
In future, we would like to implement these mechanisms to study the avalanche process.\\

Other than glitches, another interesting phenomenon of inhomogeneous vortex line movement could be 
the generations of gravitational waves from an isolated pulsar \citep{jones2002, layek2015, pradeep}. We would 
like to explore this possibility in future following the approach of \cite{melatos2012}, and estimate the
strain amplitude associated with the gravitational waves.

\section{Conclusion}
We proposed a novel mechanism for the unpinning of superfluid vortices in the inner crust of a pulsar. It occurs 
through the scattering of excited neutrons with the vortex core neutrons. The excitation of neutrons are caused by the absorption of
strain energy released due to the crustquake event. We take a cubical shape region near the most probable quake site around the star's 
equatorial plane and determine the volume ($\sim 10^6~ \text {m}^3 $) where
a fraction ($\frac{\Delta_f}{E_f}$) of bulk superfluid neutrons are excited. The scattering of these exited neutrons
with the vortex core neutrons results in the unpinning of vortices from the above volume. The Crab and Vela pulsar
with $\Omega \simeq 10^2~\text{s}^{-1}$ can release about $10^{13}$ vortices as a result of local unpinning.
The size of the glitches have been estimated to lie in the range 
$\sim 10^{-8} - 10^{-7}$, and $\sim 10^{-9} - 10^{-8}$ for Crab and Vela pulsar, respectively. The glitches, though vary in size, 
have the same frequency of occurrence of about once in a year.\\

We estimated the relaxation time scale of excited neutrons through neutron-vortex scattering and the value
of $\tau_{nn} \simeq 10^{-5}~\text{s}$ justifies the absence of multiple glitches within the time interval of
a few days or months. The glitch rise time $t_c \sim (0.2 - 2)~\text{s}$ in our model also turns out to be consistent 
with the typical feature of the glitch profile (sudden spin-up event). At the same time, this  
common feature (i.e., small glitch rise time) of all crustquake initiated glitch models makes it difficult to choose 
one among various models \citep{ruderman1991c, link-epstein96, akbal2017}. \\

The model for unpinning proposed here has the potential to explore further by implementing the knock-on picture to study 
vortex avalanches. Also, sudden release of a large number
of vortices can have consequences on the emission of gravitational radiation. We would like to explore these in
our future work. Finally, though we have discussed the excitation of neutron superfluid initiated by crustquake, 
this proposal is very generic. The approach used here should be applicable for any other sources, which have the 
potential to excite the superfluid neutrons, or can make superfluid - normal phase transition in the inner 
crust of a pulsar. It will be interesting to look for such sources.

\section{Acknowledgements}

We would like to thank Partha Bagchi and Arpan Das for useful discussions. We thank the 
anonymous reviewer for critical comments and constructive suggestions on the previous version of this manuscript. 

\section{Data Availability}
No new data were generated or analysed in support of this research.

\bibliographystyle{mn2e}
\bibliography{yadav-layek-final}

\begin{thebibliography}{}

\bibitem[\protect\citeauthoryear{Aasi, Abadie \& et. al.}{Aasi
  et~al.}{2013}]{aasi2013}
Aasi J.,  Abadie J.,    et. al. 2013, Phys. Rev. D, 87, 042001

\bibitem[\protect\citeauthoryear{Aasi, Abadie \& et. al.}{Aasi
  et~al.}{2014}]{aasi2014}
Aasi J.,  Abadie J.,    et. al. 2014, The Astrophysical Journal, 785, 119

\bibitem[\protect\citeauthoryear{Abadie, P. \& et. al.}{Abadie
  et~al.}{2011}]{vela2006}
Abadie J.,  P. A.~B.,    et. al. 2011, Phys. Rev. D, 83, 042001

\bibitem[\protect\citeauthoryear{Abbott, Abbott, Abraham \& et. al.}{Abbott
  et~al.}{2020}]{ligo2020}
Abbott R.,  Abbott T.~D.,  Abraham S.,    et. al. 2020, arXiv, p. 2007.14251

\bibitem[\protect\citeauthoryear{Akbal \& Alpar}{Akbal \&
  Alpar}{2018}]{akbal2017}
Akbal O.,  Alpar M.~A.,  2018, Monthly Notices of the Royal Astronomical
  Society, 473, 621

\bibitem[\protect\citeauthoryear{Alpar, Cheng \& Pines}{Alpar
  et~al.}{1989}]{alpar1989vortex}
Alpar M.,  Cheng K.,    Pines D.,  1989, The Astrophysical Journal, 346, 823

\bibitem[\protect\citeauthoryear{Alpar, Pines, Anderson \& Shaham}{Alpar
  et~al.}{1984}]{alpar1984}
Alpar M.~A.,  Pines D.,  Anderson P.~W.,    Shaham J.,  1984, The Astrophysical
  Journal, 276, 325

\bibitem[\protect\citeauthoryear{Anderson \& Itoh}{Anderson \&
  Itoh}{1975}]{anderson75}
Anderson P.,  Itoh N.,  1975, Nature, 256, 25

\bibitem[\protect\citeauthoryear{Andersson, Glampedakis, Ho \&
  Espinoza}{Andersson et~al.}{2012}]{andersson2012}
Andersson N.,  Glampedakis K.,  Ho W. C.~G.,    Espinoza C.~M.,  2012, Phys.
  Rev. Lett., 109, 241103

\bibitem[\protect\citeauthoryear{Ashton, Lasky, Graber \& Palfreyman}{Ashton
  et~al.}{2019}]{ashton2019}
Ashton G.,  Lasky P.~D.,  Graber V.,    Palfreyman J.,  2019, Nature Astronomy,
  3, 1143

\bibitem[\protect\citeauthoryear{Bagchi, Das, Layek \& Srivastava}{Bagchi
  et~al.}{2015}]{layek2015}
Bagchi P.,  Das A.,  Layek B.,    Srivastava A.~M.,  2015, Physics Letters B,
  747, 120

\bibitem[\protect\citeauthoryear{Baiko \& Chugunov}{Baiko \&
  Chugunov}{2018}]{baiko2018}
Baiko D.~A.,  Chugunov A.~I.,  2018, Monthly Notices of the Royal Astronomical
  Society, 480, 5511

\bibitem[\protect\citeauthoryear{Basu, Joshi, Krishnakumar, Bhattacharya,
  Nandi, Bandhopadhay, Char \& Manoharan}{Basu et~al.}{2020}]{basu2020}
Basu A.,  Joshi B.~C.,  Krishnakumar M.~A.,  Bhattacharya D.,  Nandi R.,
  Bandhopadhay D.,  Char P.,    Manoharan P.~K.,  2020, Monthly Notices of the
  Royal Astronomical Society, 491, 3182

\bibitem[\protect\citeauthoryear{Baym \& Pines}{Baym \& Pines}{1971}]{baym1971}
Baym G.,  Pines D.,  1971, Annals of Physics, 66, 816

\bibitem[\protect\citeauthoryear{Bhattacharyya}{Bhattacharyya}{2020}]{sudip2020}
Bhattacharyya S.,  2020, Monthly Notices of the Royal Astronomical Society

\bibitem[\protect\citeauthoryear{Chamel \& Haensel}{Chamel \&
  Haensel}{2008}]{chamel2008}
Chamel N.,  Haensel P.,  2008, Living Reviews in Relativity, 11

\bibitem[\protect\citeauthoryear{Chugunov \& Horowitz}{Chugunov \&
  Horowitz}{2010}]{chugunov2010}
Chugunov A.~I.,  Horowitz C.~J.,  2010, Monthly Notices of the Royal
  Astronomical Society, 407, L54

\bibitem[\protect\citeauthoryear{Cordes, Downs \& Krause-Polstorff}{Cordes
  et~al.}{1988}]{cordes1988}
Cordes J.~M.,  Downs G.~S.,    Krause-Polstorff J.,  1988, The Astrophysical
  Journal, 330, 847

\bibitem[\protect\citeauthoryear{Eichler \& Shaisultanov}{Eichler \&
  Shaisultanov}{2010}]{eichler2010}
Eichler D.,  Shaisultanov R.,  2010, The Astrophysical Journal Letters, 715,
  L142

\bibitem[\protect\citeauthoryear{Espinoza, Lyne, Stappers \& Kramer}{Espinoza
  et~al.}{2011}]{espinoza}
Espinoza C.~M.,  Lyne A.~G.,  Stappers B.~W.,    Kramer M.,  2011, Monthly
  Notices of the Royal Astronomical Society, 414, 1679

\bibitem[\protect\citeauthoryear{Feibelman}{Feibelman}{1971}]{feibelman1971}
Feibelman P.~J.,  1971, Phys. Rev. D, 4, 1589

\bibitem[\protect\citeauthoryear{Franco, Link \& Epstein}{Franco
  et~al.}{2000}]{franco2000}
Franco L.~M.,  Link B.,    Epstein R.~I.,  2000, The Astrophysical Journal,
  543, 987

\bibitem[\protect\citeauthoryear{Gügercinoğlu \& Alpar}{Gügercinoğlu \&
  Alpar}{2020}]{alpar2020}
Gügercinoğlu E.,  Alpar M.~A.,  2020, Monthly Notices of the Royal
  Astronomical Society, 496, 2506–2515

\bibitem[\protect\citeauthoryear{Haskell, Jones \& Andersson}{Haskell
  et~al.}{2006}]{haskell2006mountain}
Haskell B.,  Jones D.~I.,    Andersson N.,  2006, Monthly Notices of the Royal
  Astronomical Society, 373, 1423

\bibitem[\protect\citeauthoryear{Haskell \& Melatos}{Haskell \&
  Melatos}{2015}]{haskell}
Haskell B.,  Melatos A.,  2015, International Journal of Modern Physics D, 24,
  1530008

\bibitem[\protect\citeauthoryear{Haskell, Priymak, Patruno, Oppenoorth, Melatos
  \& Lasky}{Haskell et~al.}{2015}]{haskell2015}
Haskell B.,  Priymak M.,  Patruno A.,  Oppenoorth M.,  Melatos A.,    Lasky
  P.~D.,  2015, Monthly Notices of the Royal Astronomical Society, 450, 2393

\bibitem[\protect\citeauthoryear{Horowitz \& Kadau}{Horowitz \&
  Kadau}{2009}]{horowitz2009}
Horowitz C.~J.,  Kadau K.,  2009, Phys. Rev. Lett., 102, 191102

\bibitem[\protect\citeauthoryear{Jankowski, Bailes, Barr, Bateman, Bhandari,
  Briggs, Caleb, Campbell-Wilson, Flynn, Green, Hunstead, Jameson, Keane, Ravi,
  Krishnan \& van Straten}{Jankowski et~al.}{2015}]{jankowski2015}
Jankowski F.,  Bailes M.,  Barr E.,  Bateman T.,  Bhandari S.,  Briggs F.,
  Caleb M.,  Campbell-Wilson D.,  Flynn C.,  Green A.,  Hunstead R.,  Jameson
  A.,  Keane E.,  Ravi V.,  Krishnan V.~V.,    van Straten W.,  2015, The
  Astronomer's Telegram, 6903, 1

\bibitem[\protect\citeauthoryear{Jones}{Jones}{2002}]{jones2002}
Jones D.~I.,  2002, Class.Quant.Grav., 19, 1255

\bibitem[\protect\citeauthoryear{Jones \& Andersson}{Jones \&
  Andersson}{2001}]{jones2001}
Jones D.~I.,  Andersson N.,  2001, Monthly Notices of the Royal Astronomical
  Society, 324, 811

\bibitem[\protect\citeauthoryear{Keer \& Jones}{Keer \& Jones}{2015}]{keer2015}
Keer L.,  Jones D.~I.,  2015, Monthly Notices of the Royal Astronomical
  Society, 446, 865

\bibitem[\protect\citeauthoryear{Lander, Andersson, Antonopoulou \&
  Watts}{Lander et~al.}{2015}]{lander2015}
Lander S.~K.,  Andersson N.,  Antonopoulou D.,    Watts A.~L.,  2015, Monthly
  Notices of the Royal Astronomical Society, 449, 2047

\bibitem[\protect\citeauthoryear{Layek \& Yadav}{Layek \&
  Yadav}{2020}]{pradeep}
Layek B.,  Yadav P.,  2020, Journal of Astrophysics and Astronomy, 41

\bibitem[\protect\citeauthoryear{Link \& Epstein}{Link \&
  Epstein}{1996}]{link-epstein96}
Link B.,  Epstein R.,  1996, The Astrophysical Journal, 457, 844

\bibitem[\protect\citeauthoryear{Link \& Epstein}{Link \&
  Epstein}{1991}]{link1991}
Link B.~K.,  Epstein R.~I.,  1991, The Astrophysical Journal, 373, 592

\bibitem[\protect\citeauthoryear{McCulloch, Hamilton, McConnell \&
  King}{McCulloch et~al.}{1990}]{mcculloch1990}
McCulloch P.~M.,  Hamilton P.~A.,  McConnell D.,    King E.~A.,  1990, Nature
  Publishing Group, 346, 822

\bibitem[\protect\citeauthoryear{Melatos, Peralta \& Wyithe}{Melatos
  et~al.}{2008}]{melatos2008}
Melatos A.,  Peralta C.,    Wyithe J. S.~B.,  2008, The Astrophysical Journal,
  672, 1103

\bibitem[\protect\citeauthoryear{Pastore, Baroni \& Losa}{Pastore
  et~al.}{2011}]{pastore2011}
Pastore A.,  Baroni S.,    Losa C.,  2011, Phys. Rev. C, 84, 065807

\bibitem[\protect\citeauthoryear{Piekarewicz, Fattoyev \& Horowitz}{Piekarewicz
  et~al.}{2014}]{piekarewicz2014}
Piekarewicz J.,  Fattoyev F.~J.,    Horowitz C.~J.,  2014, Phys. Rev. C, 90,
  015803

\bibitem[\protect\citeauthoryear{Pizzochero}{Pizzochero}{2011}]{pizzochero2011}
Pizzochero P.~M.,  2011, The Astrophysical Journal, 743, L20

\bibitem[\protect\citeauthoryear{Ruderman}{Ruderman}{1969}]{ruderman69}
Ruderman M.,  1969, Nature Publishing Group, 223, 597

\bibitem[\protect\citeauthoryear{Ruderman}{Ruderman}{1976}]{ruderman1976}
Ruderman M.,  1976, The Astrophysical Journal, 203, 213

\bibitem[\protect\citeauthoryear{Ruderman}{Ruderman}{1991}]{ruderman1991c}
Ruderman M.,  1991, The Astrophysical Journal, 382, 587

\bibitem[\protect\citeauthoryear{Sedrakian}{Sedrakian}{1995}]{sedrakian1995}
Sedrakian A.~D.,  1995, Monthly Notices of the Royal Astronomical Society, 277,
  225

\bibitem[\protect\citeauthoryear{Sinha \& Sedrakian}{Sinha \&
  Sedrakian}{2015}]{sinha2015}
Sinha M.,  Sedrakian A.,  2015, Phys. Rev. C, 91, 035805

\bibitem[\protect\citeauthoryear{Smoluchowski \& Welch}{Smoluchowski \&
  Welch}{1970}]{smolu70}
Smoluchowski R.,  Welch D.~O.,  1970, Phys. Rev. Lett., 24, 1191

\bibitem[\protect\citeauthoryear{Thompson \& Duncan}{Thompson \&
  Duncan}{1995}]{thompson1995}
Thompson C.,  Duncan R.~C.,  1995, Monthly Notices of the Royal Astronomical
  Society, 275, 255

\bibitem[\protect\citeauthoryear{Warszawski \& Melatos}{Warszawski \&
  Melatos}{2012a}]{melatos2012}
Warszawski L.,  Melatos A.,  2012a, Monthly Notices of the Royal Astronomical
  Society, 423, 2058

\bibitem[\protect\citeauthoryear{Warszawski \& Melatos}{Warszawski \&
  Melatos}{2012b}]{warzawski2013}
Warszawski L.,  Melatos A.,  2012b, Monthly Notices of the Royal Astronomical
  Society, 428, 1911

\bibitem[\protect\citeauthoryear{Warszawski, Melatos \& Berloff}{Warszawski
  et~al.}{2012}]{warszawski2012}
Warszawski L.,  Melatos A.,    Berloff N.~G.,  2012, Phys. Rev. B, 85, 104503

\end{thebibliography}
\end{document}